\documentclass[twocolumn]{aastex62}
\usepackage{aas_macros}
\usepackage{float}
\usepackage{amsmath,amssymb}
\usepackage{natbib} 
\usepackage{graphicx} 
\usepackage{color}
\usepackage{verbatim}
\usepackage{multirow}

\newcommand{\figdir}{.} 
\newcommand{\proptosim}{\mathrel{\vcenter{
 \offinterlineskip\halign{\hfil$##$\cr
 \propto\cr\noalign{\kern2pt}\sim\cr\noalign{\kern-2pt}}}}}

\newcommand{\unit}[1]{{\rm\, #1}}

\newcommand{\mean}[1]{\langle #1\rangle}

\renewcommand{\min}{\mathrm{min}}
\renewcommand{\max}{\mathrm{max}}
\newcommand{\au}{\unit{au}}
\newcommand{\cm}{\unit{cm}}
\newcommand{\m}{\unit{m}}
\newcommand{\g}{\unit{g}}
\renewcommand{\Bar}{\unit{bar}}
\newcommand{\mBar}{\unit{mbar}}

\newcommand{\K}{\unit{K}} 
\newcommand{\km}{\unit{km}}

\renewcommand{\micron}{\unit{\mu m}}

\newcommand{\erg}{\unit{erg}}
\newcommand{\eV}{\unit{eV}}
\newcommand{\keV}{\unit{keV}}
\newcommand{\s}{\mathrm{s}}
\newcommand{\yr}{\mathrm{yr}}

\newcommand{\ang}{\ensuremath{\mathrm{\AA}}}

\newcommand{\lya}{\text{Ly}\ensuremath{\alpha~}}

\newcommand{\kb}{k_\mathrm{B}}
\newcommand{\sigb}{\sigma_\textsc{sb}}
\renewcommand{\d}{\mathrm{d}}
\newcommand{\e}{\mathrm{e}}

\newcommand{\p}{\mathrm{p}}     

\newcommand{\eff}{\mathrm{eff}}

\newcommand{\crit}{\mathrm{crit}}

\newcommand{\rcb}{\mathrm{rcb}} 
\newcommand{\eq}{\mathrm{eq}}   
\renewcommand{\H}{\mathrm{H}}     

\newcommand{\ratio}[2]{\ensuremath{\left(\dfrac{#1}{#2}
      \right)}}
\begin{document}

\title{Dusty outflows in planetary atmospheres:
  Understanding ''super-puffs'' and transmission spectra of
  sub-Neptunes}

\author{Lile Wang$^{1,2}$, Fei Dai$^{2,3}$}

\footnotetext[1]{Center for Computational Astrophysics,
  Flatiron Institute, \\ New York, NY 10010;
  lwang@flatironinstitute.org} \footnotetext[2]{Princeton
  University Observatory, Princeton, NJ 08544}
\footnotetext[3]{Department of Physics and Kavli Institute
  for Astrophysics \\
  and Space Research, Massachusetts Institute of Technology,
  \\
   Cambridge, MA 02139}

\begin{abstract}
  ``Super-puffs'' are planets with anomalously low mean
  densities ($\lesssim 10^{-1}~\g~\cm^{-3}$). With a low
  surface gravity, the extended atmosphere is susceptible to
  extreme hydrodynamic mass loss (``boil off'') on a
  timescale much shorter than the system's age. Even more
  puzzling, super-puffs are estimated to have a scale height
  of $\sim 3000~\km$, yet recent observations revealed
  completely flat transmission spectra for Kepler 51b and
  51d. We investigate a new scenario that explains both
  observations: non-static outflowing
  ($\dot{M}\gtrsim 10^{-10}~M_\oplus~\yr^{-1}$) atmospheres
  that carry very small dust grains ($\sim 10~\ang$ in size,
  $\sim 10^{-2}$ in mass fraction) to high altitudes
  ($\lesssim 10^{-6}~\Bar$).  Dust at high altitudes
  inflates the observed transit radius of the planet while
  flattens the transmission spectra.Previous static
  atmospheric models struggles to achieve cloud elevation
  and production of photochemical haze at such high
  altitudes. We propose to test this scenario by extending
  the wavelength coverage of transmission spectra. If true,
  dusty atmospheric outflows may affect many young
  ($\lesssim 10^9~\yr$), low mass ($\lesssim 10~M_\oplus$)
  exoplanets, thereby limit our ability to study the
  atmospheric composition in transmission, and inflate the
  observed transit radius of a planet hence obscure the
  underlying mass-radius relationship.
\end{abstract}


\section{Introduction}
\label{sec:intro}

``Super-puffs'' are planets that have sub-Neptune masses
($\lesssim 5~M_\oplus$) but gas-giant transit radii
($\gtrsim 5~R_\oplus$), and thus extremely low mean
densities ($<10^{-1}~\g~\cm^{-3}$) and large scale heights
($\sim 3000~\km$). A prime example is Kepler 51b, which has
a $\sim 7 R_\oplus$ transit radius but a mass of only
$\sim 2.1~M_\oplus$ \citep[consolidated by independent
transit timing variation analyses of several groups
e.g. Roberts et al. in prep; ][M14 hereafter]
{2014ApJ...783...53M}. The ensemble of discovered
super-puffs include Kepler 51c, 51d; Kepler 79d, 79e
\citep{2014ApJ...785...15J}; and Kepler 87c
\citep{2014A&A...561A.103O}. In this letter we concentrate
our discussions on the well-studied Kepler 51b unless
specially noted.

Recent works (\citealt{2017ApJ...847...29O,
  2018ApJ...860..175W}, WD18 hereafter) suggest that
hydrodynamic and photoevaporative loss of atmospheres might
be a ubiquitous effect responsible of the observed bimodal
radius distribution of close-in sub-Neptune planets
\citep{2017AJ....154..109F}. Given their low surface
gravity, super-puffs are expected to have excessive
hydrodynamic mass-loss even without stellar high energy
radiation \citep[``boil-off'', see also][]
{2016ApJ...817..107O}, and should disperse on a timescale of
$\sim 10^3~\yr$ (\S\ref{sec:intro-atm}), much shorter than
the system's age ($\sim 0.3~{\rm Gyr}$ for Kepler 51 from
gyrochronology; M14). {Similarly,
  \citet{2016MNRAS.461L..62L} noted that CoRoT-24b must also
  have high-altitude aerosols to increase the apparent
  transit radius, thereby lowering the implied mass loss
  rate. However, they were agnostic of how aerosols could
  form or be lifted to such hight altitudes.}

Super-puffs, with their large scale heights, are considered
ideal targets for transmission spectroscopy. However, the
HST WFC3 observation of Kepler 51b and 51d yield flat
transmission spectra in the near-infrared (Roberts et al. in
prep). This is reminiscent of the flat spectrum of GJ1214b
\citep{2014Natur.505...69K}. If cloud/haze are invoked to
mute the absorption features, they have to be advected to or
produced at such a high altitude that current models would
struggle with (\S\ref{sec:intro-dusts}). We hereby consider
a non-static atmosphere characterized by a slow hydrodynamic
outflow ($\gtrsim 10^{-10}~M_\oplus~\yr^{-1}$), producing a
relatively small mass-loss over the age of Kepler 51. Dust
grains can be carried to much higher altitude in this
outflow, increasing the observed transit radius to
$\sim 7~R_\oplus$ while muting signatures of other species
in the atmosphere.

\section{Basic ideas}
\label{sec:intro-motiv}


\subsection{Isothermal atmosphere: Inevitable escape}
\label{sec:intro-atm}

Generally a planetary atmosphere can be divided into a
convective isentropic interior and a radiation-dominated,
approximately isothermal exterior
\citep{2006ApJ...648..666R, 2016ApJ...817..107O,
  2016ApJ...825...29G}. In the isothermal layer, hydrostatic
density and pressure profiles are given by,
\begin{equation}
  \label{eq:intro-iso-atm}
  \begin{split}
    & p = p_\p \exp\left[ \beta_\p \left( \dfrac{r_\p}{r} -
      1 \right) \right] \geq p_\infty = p_\p\e^{-\beta_\p}\ ;
    \\
    & \rho =\dfrac{p\mu}{\kb T_\eq}\ ;\quad
    \beta_\p \equiv \dfrac{G M_\p \mu} {r_\p \kb T_\eq}\ ,
  \end{split}  
\end{equation}
where the subscripts ``p'' and ``$\infty$'' denote the
quantities at the planetary radius and infinite radius
respectively, $G$ is the gravitational constant, $\kb$ is
the Boltzmann constant, $M_\p$ is the planetary mass (core
and atmosphere combined), $\mu$ is the (dimensional) mean
molecular mass, and
$T_\eq \simeq 886~\K\ \left( {L_*}/{L_\odot} \right)^{1/4}
\left({a}/{0.1~\au} \right)^{-1/2}$ is the equilibrium
temperature at planetary orbit radius $a$ and host star
luminosity $L_*$. The dimensionless parameter $\beta_\p$ is
also called the ``restricted Jeans parameter''
\citep[e.g.][]{2017A&A...598A..90F, 2017MNRAS.466.1868C}.
We also remind the reader that $p_\infty$ in
eq.~\ref{eq:intro-iso-atm} serves as a confining term
preventing the isothermal atmosphere from a spontaneous
outflow.
{If one naively assumes a clear atmosphere (free of
  cloud/haze) of solar abundance, $p_\p\sim 20-100~\mBar$ is
  required at the observed transit radius
  \citep[e.g.][]{2014ApJ...792....1L,
    2016MNRAS.461L..62L}. For Kepler 51b, this leads to
  $\beta_\p \simeq 9.8$ and $p_\infty\sim 10^{-6}~\Bar$
  using eq.~\eqref{eq:intro-iso-atm} } Such $p_\infty$ is a
few orders of magnitude greater than any plausible sun-like
stellar wind total pressure
\citep[]{2009ApJ...693...23M}. Unconfined atmospheres
hydrodynamically lose mass at
$\dot{M} \sim \min\{ \dot{M}_{\rm Parker},\ \dot{M}_{\rm
  rad} \}$, where \citep[e.g.][]{1958ApJ...128..664P},
\begin{equation}
  \label{eq:intro-mdot-parker}
  \begin{split}
    & \dot{M}_{\rm Parker} \sim 4\pi r_\s^2 c_\s \rho_\p
    \exp \left( \dfrac{3}{2} - \dfrac{2r_s}{r_\p} \right)\ ,
    \\
    & \dot{M}_{\rm rad} \sim \left( \dfrac{L_*}{4\pi a^2}
    \right) \pi r_\p^2 \left(\dfrac{c_s^2}{2}\right)^{-1}\ .
  \end{split}
\end{equation}
Here $c_s = (\kb T_\eq/ \mu)^{1/2}$ is the isothermal sound
speed and $r_s = G M_\p / (2c_\s^2) $ is the sonic
radius. We find $\dot{M}\sim 10^{-3}~M_\oplus~\yr^{-1}$ with
$p_\p \sim 10~\mBar$ for Kepler 51b, dispersing the
atmosphere in $\lesssim 10^3~\yr$---much shorter than the
estimated age of the system ($\sim 0.3~{\rm Gyr}$), which in
turn questions the earlier assumption of ``clear''
atmosphere.

\subsection{Dusts in the Atmospheres}
\label{sec:intro-dusts}

Aerosols, which could consist of dusts and liquid droplets,
could dramatically increase the opacity of gas.
The enhanced opacity lowers the required pressure at the
apparent planet radius $p_\p$ by several orders of
magnitude, giving rise to a much slower outflow.
However, maintaining aerosol particles at a radius as high
as $7~R_\oplus$ over Kepler 51b is difficult in a static
atmosphere. In-situ formation of dusts (for clouds/haze)
demands rather high gas density; photochemical calculations
reveal that dust formation is very inefficient below
$p\sim 10^{-7}-10^{-6}~\Bar$ \citep[][]
{2012ApJ...756..172M, 2013ApJ...775...33M,
  2013ApJ...775...80F, 2018ApJ...853....7K}. 
Aerosols are also subject to planetary gravity; dust grains
with radius $r_\d$ precipitate at terminal velocity
$v_{\rm term}$ and timescale $\tau_{\rm prec}$
\citep{1965MNRAS.130...63B,DraineBook},
\begin{equation}
  \label{eq:intro-vterm}
  \begin{split}
    v_{\rm term} & \sim 3~\m~\s^{-1} \times
    \ratio{\mu}{m_\H}^{-1/2} \ratio{M_\p}{M_\oplus}
    \ratio{r}{R_\oplus}^{-2}
    \\
    & \times\ratio{r_\d}{10~\ang}\ratio{\rho / m_\H}
    {10^{12}~\cm^{-3}} \ratio{T}{10^3~\K}^{-1/2}\ ;
    \\
    \tau_{\rm prec,p} & \equiv \dfrac{r_\p}{v_{\rm term,p}}
    \sim 10^{-1}~\yr \ratio{r_\d}{10~\ang}^{-1}
    \ratio{\rho_\p / m_\H}{10^{12}~\cm^{-3}}\ .
  \end{split}
\end{equation}
{The eddy diffusion coeffient required to lift
  $r_\d = 10~\ang$ dusts to $\sim 7~R_\oplus$ is at least
  $K_{zz}\sim 10^{11}-10^{12}~\cm^2~\s^{-1}$, which is
  significantly greater than the values observed on the
  Earth \citep{2015JGRA..120.3097P} and modeled on
  exoplanets \citep{2013ApJ...775...33M}. } Even if dust
formation at high altitudes were sufficient to compensate
dust precipitation, in a static atmosphere with the
eq.~\eqref{eq:intro-iso-atm} density profile, heavy elements
in this layer are rapidly depleted at timescale
$\lesssim (m_{\rm metal}/m_\d)\tau_{\rm prec}$ [here
$(m_{\rm metal}/m_\d)$ is the atmospheric mass ratio of
metal elements to dusts].

We thus consider non-static atmospheres in which aerosols
are co-moving with outflows. The critical mass-loss rate,
at which $v_r=v_{\rm term}$ (note that this Equation does
{\it not} depend on $r$; see also WD18)
\begin{equation}
  \label{eq:mdot-crit}
  \begin{split}
    \dot{M}_\crit & \equiv 4\pi r^2 \rho v_{\rm term}
    \simeq  2\times 10^{-11}~M_\oplus~\yr^{-1}
    \ratio{M_c}{M_\oplus}  \\
    & \times \ratio{r_\d}{10~\ang}
    \ratio{T}{10^3~\K}^{-1/2} \ratio{\mu}{m_\H}^{1/2}\ .
  \end{split}
\end{equation}
Whenever $\dot{M} \gg \dot{M}_\crit$, dusts experience
neglibible precipitation, and can be considered as co-moving
with gas. $\dot{M}$ must also satisfy
$\dot{M} < \dot{M}_\max\sim (M_{\rm atm} / \tau_\p)$, where
$M_{\rm atm}$ is the total mass of atmosphere and $\tau_\p$
is the planet's age (approximated by the host star's age
$\tau_*$; for Kepler 51b,
$\dot{M}_\max \sim
10^{-9}~M_\oplus~\yr^{-1}$). {Dusts of
  $\sim 10~\ang$ sizes should be abundantly produced by
  geological activities, while laboratory experiments
  \citep{2018NatAs...2..973Z} show that gas-phase formation
  of tiny graphites and polycyclic aromatic hydrocarbon
  (PAH) can also be very efficient even at relatively low
  temperatures and UV intensities. Meanwhile, the
  temperature throughout most of the internal atmosphere
  (\S\ref{sec:model-isentropic}) is higher than dust
  sublimation temperature ($\sim 1500~\K$), preventing tiny
  grains from fast coagulating: larger grains fall back to
  the internal atmosphere and are broken into gaseous
  species.}

\subsection{Effective transit radii}
\label{sec:intro-tran-rad}

High-altitude aerosols lead to extra extinction on stellar
light from the observer's view, thus effectively increases
the planet trasiting radii. To ease later discussion, we
define the effective transit radius:
\begin{equation}
  \label{eq:r-eff}
  \mean{r_\eff} \simeq \left\{\dfrac{1}{\pi}
    \int_0^{R_\H}\d b\ 2\pi b \left[1-\e^{-\tau(b)}\right]
  \right\}^{1/2}\ ,
\end{equation}
where $\tau(b)$ is the optical depth along the line-of-sight
(LoS) at impact parameter $b$ relative to the planet
geometric center. The upper limit of the integral is $R_\H$
(the planet's Hill radius) where the assumption of
excluding host star gravitation likely breaks down.
We estimate the optical depth by
$\tau(b) \simeq \Sigma(b) X_\d \sigma_{\rm d, ext}$, where
$\Sigma(b)$ is the column density along the LoS,
$X_\d \simeq n_\d / (\rho / m_\H)$ is the number fraction of
dust particles relative to hydrogen nuclei, and
$\sigma_{\rm d, ext}$ is the extinction cross section of a
single dust particle. At optical and infrared (IR)
wavelengths $0.2 \lesssim (\lambda/\micron) \lesssim 2$, the
extinction cross section of very small grains is well
approximately given by a smooth power-law function,
\begin{equation}
  \label{eq:sig-ext}
  \sigma_{\rm \d,ext} \simeq \sigma_{-16}\times
  10^{-16}~\cm^{2} \ratio{r_\d}{10~\ang}^3
  \ratio{\lambda}{\micron}^{-\delta}\ ,
\end{equation}
where $(\sigma_{-16},\delta) \simeq (0.92,1.55)$ for
graphites, and $(0.11,0.93)$ for silicates
\citep{1993ApJ...414..632D}. {PAH grains at
  $r_\d\sim 10~\ang$ have an absorption edge at
  $\lambda\sim 1~\micron$, and are optically similar to
  graphites at shorter wavelengths
  \citep{2001ApJ...554..778L}.
} For simplicity we assume that all aerosols consist of
graphite dusts. The dust-to-gas mass ratio corresponding to
number ratio $X_\d$ is, assuming hydrogen atmosphere,
\begin{equation}
  \label{eq:mdust-mgas}
  \dfrac{m_\d}{m_{\rm gas}} \simeq\ X_\d
  \ratio{m_{\rm C}N_{\rm C,dust}}{m_\H} \simeq
  0.56~\ratio{r_\d}{10~\ang}^3 \ratio{X_\d}{10^{-4}}\ ,
\end{equation}
where $N_{\rm C, dust}\simeq 470~(r_\d/10~\ang)^3$ is the
number of carbon atoms per dust grain.

\section{Detailed Modeling}
\label{sec:outflow}


\subsection{Isentropic interior}
\label{sec:model-isentropic}

Although all interesting atmospheric dynamics take place in
the radiative exterior, hydrodynamic structures of the
convective interior should still be consistently calculated
by solving,
\begin{equation}
  \label{eq:conv-ode}
   \dfrac{\d M_{\rm a}}{\d r} = 4\pi r^2 \rho\ ,\ 
   \dfrac{\d p}{\d r} = -\dfrac{G (M_{\rm a} + M_c)
     \rho}{r^2},\ p = \kappa \rho^\gamma,
\end{equation}
where $M_{\rm a}$ is the mass of atmosphere enclosed by
radius $r$, $M_c$ is the mass of the solid planet core,
$\kappa$ is the specific entropy parameter, and $\gamma$ is
the adiabatic index (we take $\gamma = 1.4$ for the molecular
atmospheres in this letter).

The gravitation in the radiative exterior of atmosphere
depends on both $M_c$ and the total mass of isentropic
atmosphere $M_{\rm atm}$, while the self-gravity of the gas
in that layer is usually negligible. In practice, we first
pick an $M_c$ and an $M_{\rm atm}$ and obtain a model of the
external radiative atmosphere. Then, we solve
eq.~\eqref{eq:conv-ode} as a boundary value problem such
that (1) $M_{\rm a}(R_c) = 0$ ($R_c$ is the planet core
radius), and (2) $p$ and $\rho$ match the external
atmosphere profiles at the radiative-convective boundary
$r_\rcb$, which is adjusted so that
$M_{\rm a}(r_\rcb) = M_{\rm atm}$. The isentropic atmosphere
is characterized by its Kelvin-Helmholtz timescale
$\tau_{\rm kh}$ \citep[e.g.][]{2017ApJ...847...29O}.

\subsection{Dusty outflowing exterior}
\label{sec:model-outflow}


The model planet orbits the host star (for simplicity, we
round off to $M_*=M_\odot$, $L_*=0.88 L_\odot$ from M14) on
a $a=0.25~\au$ circular orbit ($T_\eq = 543~\K$).
The planet combines an $M_c = 1.7~M_\oplus$,
$R_c=1.14~R_\oplus$ solid core and an
$M_{\rm atm} = 0.4~M_\oplus$ convective atmosphere.

\subsubsection{Model 0: Isothermal Parker wind}
\label{sec:model-parker}

The first model (Model 0) that we consider is constructed
{\it analytically}.  If we assume an isothermal $T = T_\eq$,
the well-known Parker wind solution satisfies
\citep{1958ApJ...128..664P},
\begin{equation}
  \label{eq:iso-mach}
  \exp\left( -\dfrac{\mathcal{M}^2}{2} \right)
  = \varrho \exp\left( \frac{3}{2} - \frac{2}{\eta}\right)
  \ ,\ \varrho = \dfrac{1}{\mathcal{M}\eta^2}\ ,
\end{equation}
where $\mathcal{M} \equiv v_r / c_s$ is the radial Mach
number, $\eta\equiv r/r_s$ is the dimensionless radius
normalized by the sonic radius $r_s$, and
$\varrho\equiv \rho/\rho_s$ is the dimensionless density
normalized by $\rho_s$ (the density at sonic radius).

\subsubsection{Consistent thermochemical simulations}
\label{sec:model-sims}

Models 1 and 2 involve full hydrodynamic simulations that
incorporate radiation and thermochemistry described in WD18.
The axisymmetric 2.5-dimensional spherical-polar mesh
centers at the planet, whose polar axis points to the host
star. It spans
$(r,\theta) \in [3~R_\oplus,400~R_\oplus]\otimes [0,\pi]$ at
resolution $256\times 128$ (radial zones are spaced
logarithmically and latitudinal zones evenly), to guarantee
that all relevant physical processes are included in the
simulation domain. 
The initial conditions obey the isothermal hydrostatics at
$T_\eq$ in eq.~\eqref{eq:intro-iso-atm}, where
$\rho_{\rm ini}(r_{\rm in})$ (the initial mass density at
the inner boundary $r_{\rm in} = 3~R_\oplus$) is the
variable parameter. Initial abundances of chemical species
are uniform across the simulation domain; they are identical
to WD18, except for the dusts. We adjust
$\rho_{\rm ini}(r_{\rm in})$ and the dust-to-gas mass ratio
$(m_\d/m_{\rm gas})$ for each simulation so that
$\mean{r_\eff} = r_\p = 7~R_\oplus$ and
$\dot{M}_\max > \dot{M} \gg \dot{M}_\crit$ in steady states.

Both models include the host star luminosity
$L(2~\eV)=0.88~L_\odot$, representing infrared and optical
radiation. Model 2 also involves high-energy photons
represented by four photon energy bins ($h\nu = 7~\eV$ for
soft FUV, $12~\eV$ for Lyman-Werner band FUV, $25~\eV$ for
EUV, and $3~\keV$ for the X-ray) at
luminosities\footnote{These high-energy luminosities are
  estimated with the recipes in \citet{2017ApJ...847...29O}
  and WD18, adopting \citet{2005ApJ...622..680R} for
  $L(t<10^8~\yr)$ and assuming
  $L/L(t<10^8~\yr) = \min\{1,(\tau_*/10^8~\yr)^{-1.5}\}$.}:
$L(7~\eV) = L(25~\eV) = L(3~\keV) = 8\times
10^{28}~\erg~\s^{-1}$, and
$L(12~\eV) = 8 \times 10^{27}~\erg~\s^{-1}$.  Rays are
parallel to the polar axis, entering the simulation domain
at the outer radial boundary with fluxes
$F(h\nu) = L(h\nu) / (4\pi a^2)$.

We include $r_\d = 10~\ang$ graphites in these two models as
a proxy of dusts of all sizes and components. Dust
temperature is estimated by the dual-temperature profile
$T_\d=\max\{T_\eq,\tilde{T}_\d\}$ (similar to
\citealt{1997ApJ...490..368C}), where $\tilde{T}_\d$ is
obtained by solving
\begin{equation}
  \label{eq:dust-temp}
  \sum_{h \nu} F(h \nu) \sigma_{\rm d, ext} (h \nu)
  = 4 \pi r_\d^2 \sigb \tilde{T}_\d^4 q(\tilde{T}_\d) \ .
\end{equation}
Here $\sigb$ is the Stefan-Boltzmann constant, and
$q(T_\d)$ is the dust emissivity.

\subsection{Results}
\label{sec:sim-results}

\begin{deluxetable}{lccccc}
  \tablecolumns{5} 
  \tabletypesize{\scriptsize}
  \tablewidth{0pt}
  \tablecaption{Properties of the representative models 
    \label{table:various-model}
  }
 \tablehead{
    \colhead{Model} \vspace{-0.25cm} &
    \colhead{$\tau_{\rm kh}$} &
    \colhead{$\rho_{\rm ini}(r_{\rm in})$} &    
    \colhead{$m_\d/m_{\rm gas}$} &
    \colhead{$\dot{M}_{10}^\dagger$}
    \\
    \colhead{  } &
    \colhead{($10^9~\yr$)} &
    \colhead{($10^{-8}\g~\cm^{-3}$)} &
    \colhead{($10^{-2}$)} &
    \colhead{ }
  }
   \startdata
   0 (Parker wind) & 2.3 & 18.9 & 1.7 & 4.0  \\
   1 (Optical \& IR)  & \multirow{2}{*}{6.2}
   & \multirow{2}{*}{0.37} & 2.6 & 5.4 \\
   2 (UV \& X-ray)  &  &  & 4.8 & 6.5 
   \enddata
   \tablecomments
   { All models have $\mean{r_\eff} = 7~R_\oplus$ at
     $\lambda=1~\micron$. 
     \\
     $\dagger$: $\dot{M}\equiv \dot{M}_{-10}\times
     10^{-10}~M_\oplus~\yr^{-1}$ } 
\end{deluxetable}

\begin{figure}
  \centering
  \includegraphics[width=3.0in, keepaspectratio]
  {\figdir/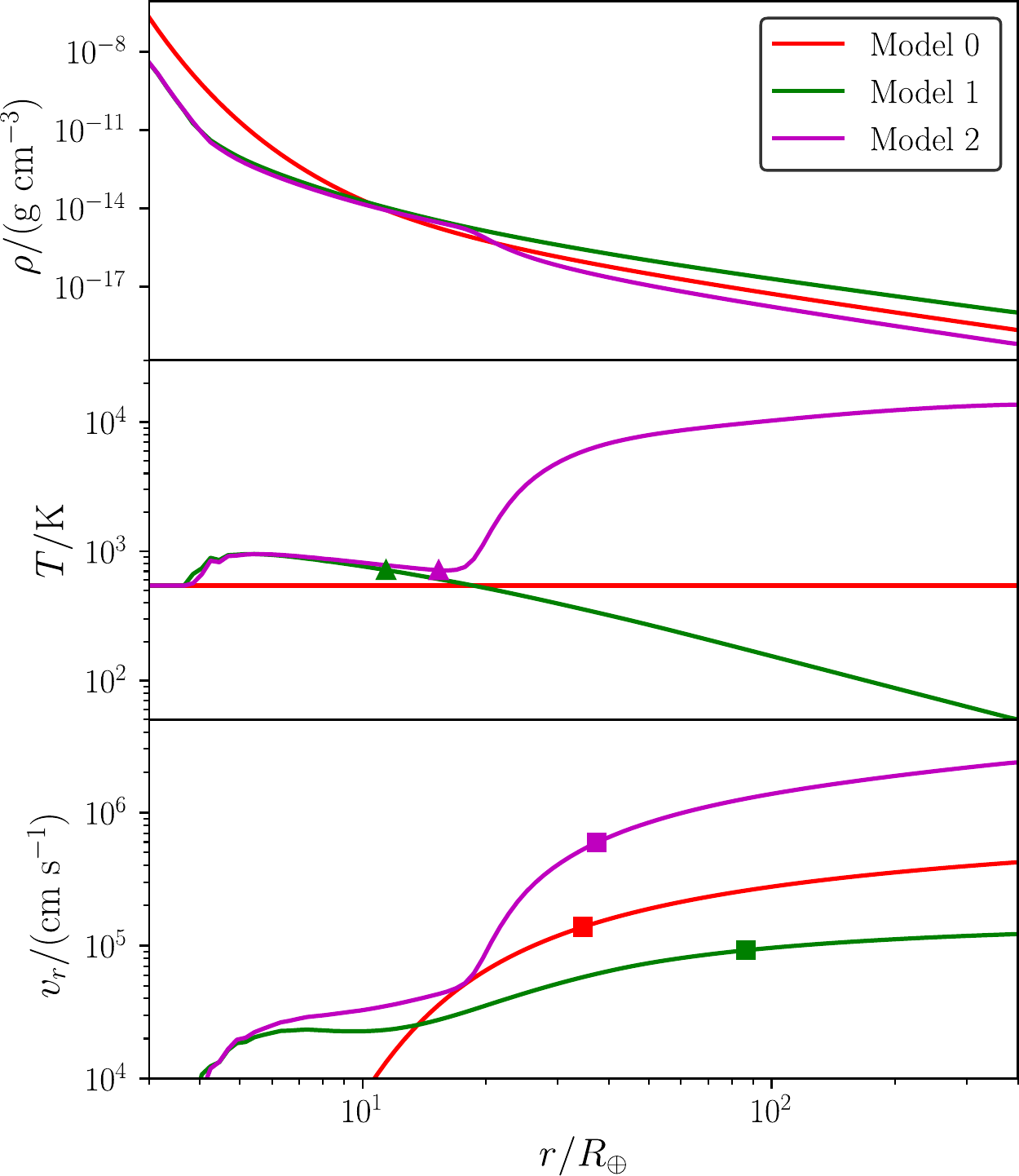}
  \caption{Hydrodynamic profiles (top: density $\rho$;
    middle: temperature $T$; bottom: radial velocity $v_r$)
    of models in Table~\ref{table:various-model} along the
    $\theta = \pi / 2$ radii. Models are distinguished by
    colors. Triangles mark the locations wher gas thermally
    decouples from dusts (above which $|T_\d-T|/T_\d>0.3$)
    for Models 1 and 2. Squares mark the radial sonic points
    ($v_r=c_s$).}
  \label{fig:hydro_profile}
\end{figure}

\begin{figure}
  \centering
  \includegraphics[width=3.0in, keepaspectratio]
  {\figdir/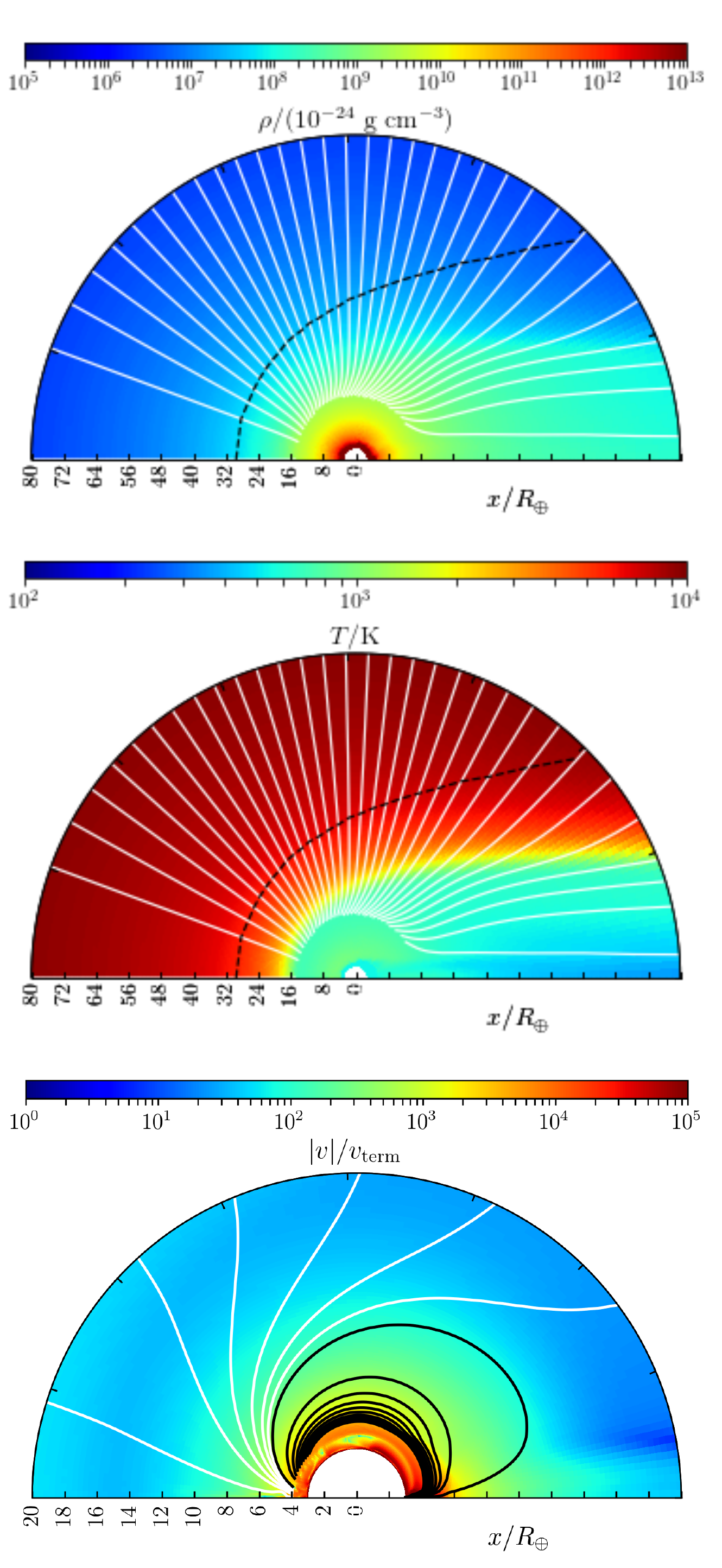}
  \caption{Meridional plot of Model 2
    (\S\ref{sec:model-sims}) in steady state, showing
    density $\rho$ (top panel), temperature $T$ (lower
    panel), and $|v|/v_{\rm term}$ (bottom panel, zoomed-in
    for the innermost $20~R_\oplus$) profiles. {\bf Top and
      middle panels} are overlaid by streamlines in white
    solid curves, separated by mass flow
    $2\times 10^{-11}~M_\oplus~\yr^{-1}$, shown only in
    regions where the total energy of fluid elements is
    positive. Sonic surface is overlaid with black dashed
    curves. {\bf Bottom panel} zooms-in for two types of
    streamliens: white curves are streamlines that
    eventually join the EUV wind and escape to infinity,
    separated by $10^{-10}~M_\oplus~\yr^{-1}$ mass flow;
    black curves are streamlines that eventually fall back,
    separated by $10^{-9}~M_\oplus~\yr^{-1}$ mass flow. Only
    the $r \geq 4~R_\oplus$ part (approximately the radius
    of $h\nu = 2~\eV$ radiation front in the day hemisphere)
    of streamlines are presented. }
  \label{fig:euv_slice}
\end{figure}

\subsubsection{Model profiles}
\label{sec:result-profiles}


Table~\ref{table:various-model} summarizes the key
properties and results of our models. All models demand
$\sim 10^{-2}$ of atmospheric mass in dusts to achieve
$\mean{r_\eff} = 7~R_\oplus$ with $M_\p = 2.1~M_\oplus$. The
gas pressure required at $r_\p$ is merely
$\sim 10^{-8}-10^{-9}~\Bar$, while the
$p=100~\mBar$ radii is much lower (compared to
\S\ref{sec:intro-atm}):
$r_{100~\mBar}\simeq 2.5~R_\oplus$ (Model 0) or
$2.2~R_\oplus$ (Models 1 and 2). Density, temperature and
radial velocity profiles along the radial ray at
$\theta = \pi/2$ (i.e. perpendicular to the direction to the
host star) of all models are presented by
Figure~\ref{fig:hydro_profile}. Figure~\ref{fig:euv_slice}
illustrates the meridional plots of density, temperature and
velocity profiels for Model 2 in steady state, which are
similar to the EUV photoevaporation models discussed in
WD18: a hot ($T>10^4~\K$), anisotropic
EUV-dominated outflow, a warm ($T\lesssim 10^3~\K$)
intermediate layer, and a ``tail'' behind the night
hemisphere.

Curiously, there are also day-night meridional motions in
Models 1 and 2. This is the consequence of dust temperature
excess: in regions accessible by $h\nu = 2~\eV$ photons,
dust temperature $T_\d\simeq 1050~\K \gg T_\eq$ due to
$q(T_\d)\ll 1$ (eq.~\ref{eq:dust-temp}), causing gas
temperature $T\gg T_\eq$ via dust-gas thermal accommodation.
Figure~\ref{fig:euv_slice} illustrate such meridional
motion, which never leaves the planetary gravity potential,
but still satisfies $|v|\gg v_{\rm term}$ hence can keep the
dusts aloft. We nonetheless choose not to over-interpret
this result: atmospheric circulation requires proper
treatment of radiative transfer, dimensionality and planet
spin to model, which are postponed to future works.

\subsubsection{Transit light curves and model consistency}
 
\begin{figure}
  \centering
  \includegraphics[width=3.1in, keepaspectratio]
  {\figdir/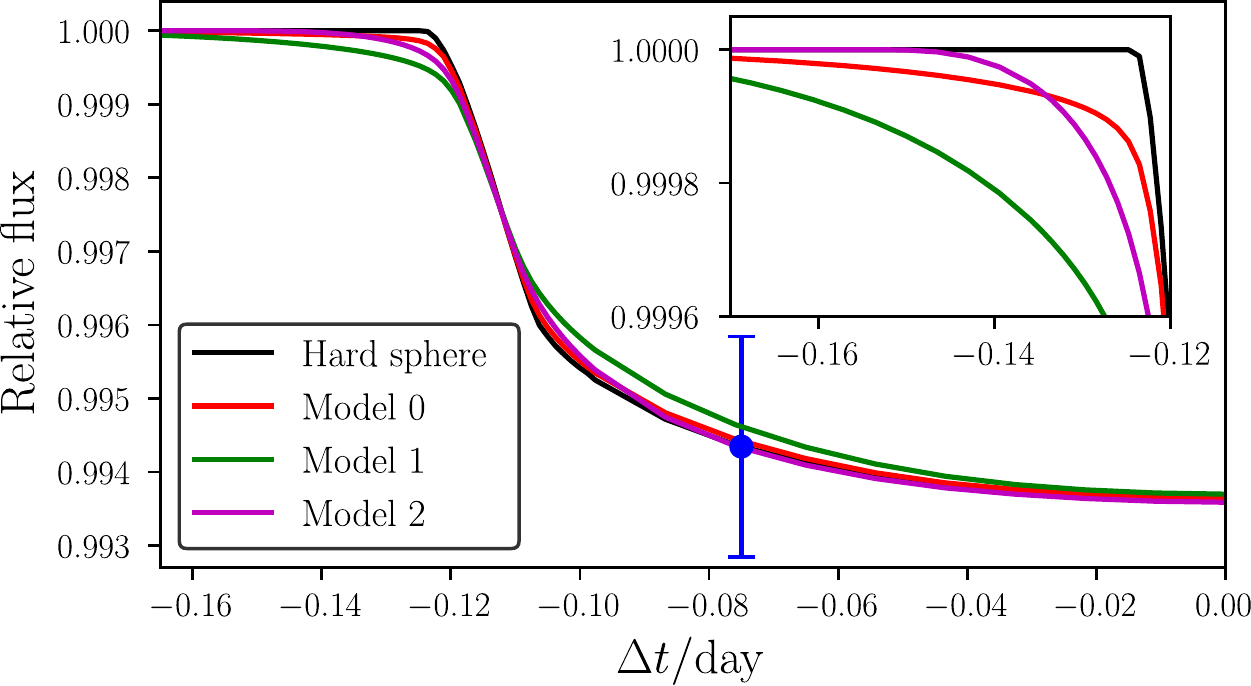}
  \caption{Transit light curves of our models (Models 0
    through 2, plus a hard sphere for reference), showing
    only the range $\Delta t < 0$ ($\Delta t > 0$ curves are
    omitted due to symmetry). An errorbar is overlaid to
    indicate the typical error of Kepler short-cadence
    measurement. The inserted panel zooms in the light
    curves near the ingress. Our dusty outflow models all
    produce a gentler ingress/egress compared to the hard
    sphere model. However, after analyzing the Kepler light
    curve, we found that these models cannot be decidedly
    ruled out due to the large observation uncertainty.}
  \label{fig:transit}
\end{figure}

Figure~\ref{fig:transit} illustrates the synthetic transit
light curves (limb darkening profile adopted from M14),
plus a simple ``hard sphere'' for reference. All models have
a extended but gentler ingress/egress than the hard-sphere.
Model 2 has a relatively sharper ingress/egress, because EUV
photons carve a cliff in density and temperature by
launching a photoevaporative wind. 
The synthetic light curve is symmetric about the
mid-transit, as $\mean{r_\eff} = 7~R_\oplus$ is still deep
in the planet's potential well. To analyze the detectability
of the difference in the light curves, we re-sample
systhetic light curves with 1-minute cadence and add a white
noise component of $1500~{\rm ppm}$ to mimic the {\it
  Kepler} observation of Kepler 51b. The resultant light
curves were analyzed with a conventional
\citet{2002ApJ...580L.171M} transit model similar to that
employed by M14. We found that more extended and gentler
ingress/egress of the synthetic light curves can be
accommodated by a combination of higher impact parameter $b$
and slightly different limb darkening coefficients than
those reported by M14. A future observation of the system
with higher photometric precision is required to distinguish
Models 0 through 2 which differs by only
$\sim 200~{\rm ppm}$.

\section{Discussion and Summary}
\label{sec:discussions}

In this letter, we showed that a dusty outflow of a
planetary atmosphere could enhance the opacity at high
altitudes, therefore successfully explains the puffy Kepler
51b, and flat transmission spectrum of super-puff
exoplanets. The dusty outflow scenario relies on the
mass-loss rate $\dot{M}$, which should stay in a proper
range ($\dot{M}_\crit \ll \dot{M} \ll \dot{M}_\max$; see
\S\ref{sec:intro-dusts}), favoring the class of young,
low-mass sub-Neptunes.  \citet{2017MNRAS.466.1868C} suggests
that $\sim 15\%$ of sub-Neptunes are too puffy and may be
currently experiencing mass loss. The mechanism is maximized
when the atmospheric dispersal timescale is similar to the
age of the system [e.g.  $\sim 0.3~{\rm Gyr}$ for Kepler 51
(M14), and $\lesssim 1~{\rm Gyr}$ for Kepler 79
\citep{2013MNRAS.436.1883W}].

\begin{figure}
  \centering
  \includegraphics[width=3.1in, keepaspectratio]
  {\figdir/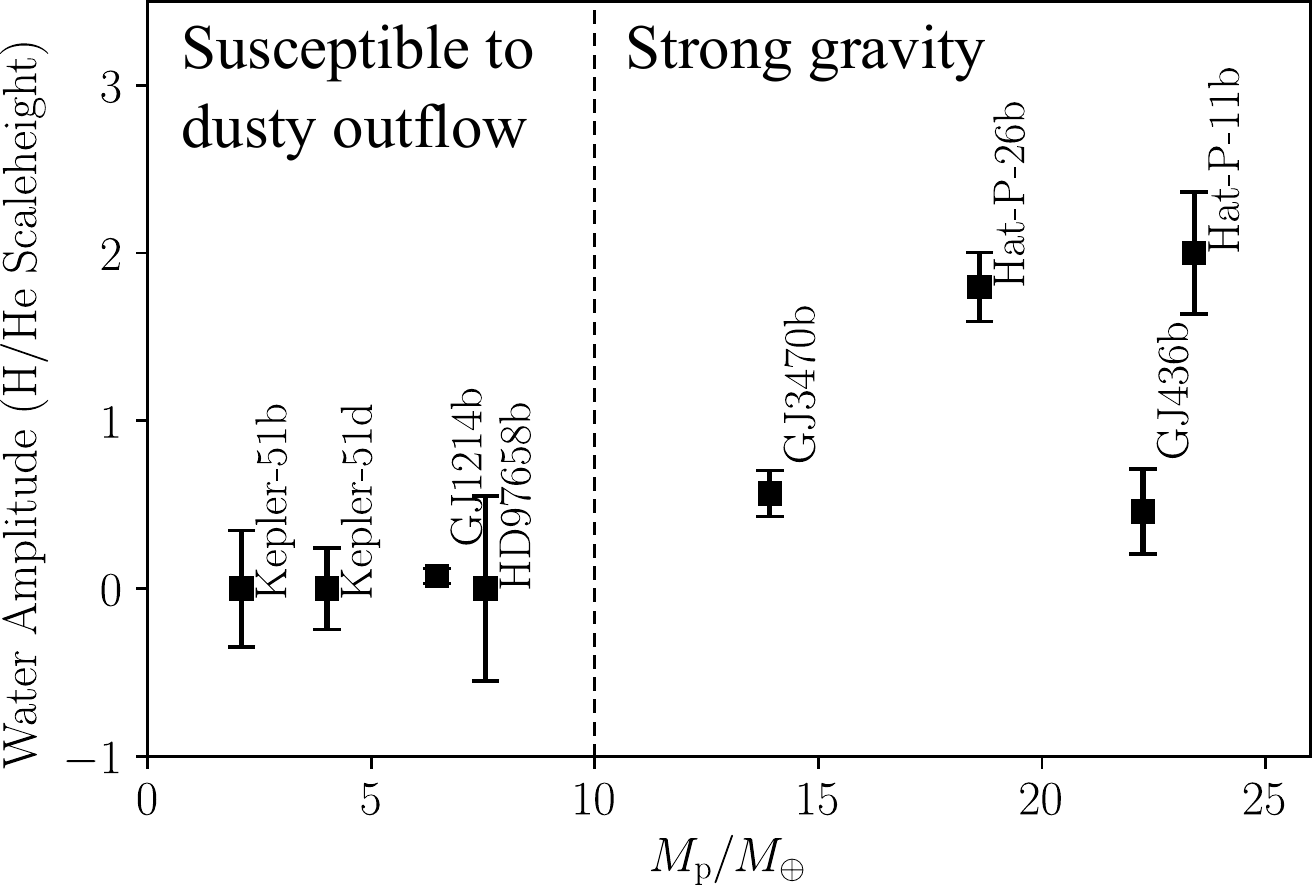}
  \caption{Spectral strenths of water feature versus planet
    mass, compiled by \citet{2017AJ....154..261C}; data of
    Kepler 51b and 51d are from Roberts et al, in
    prep. According to our earlier simulations (WD18), below
    $\sim 10~M_\oplus$ are the objects that are susceptible
    to dusty outflows, while $M_\p > 10~M_\oplus$ planets
    have too strong gravity to efficiently launch outflows.}
  \label{fig:water_feature}
\end{figure}

Dusty outflows have several implications. First, extinction
cross sections of small grains 
are smooth function of wavelengths in optical and
near-infrared \citep[see also][] {1984ApJ...285...89D,
  1993ApJ...414..632D, 2001ApJ...554..778L}. Dusts therefore
obscure the signatures of some other chemical species in
planetary atmospheres, limiting the ability of transmission
spectroscopy.  Figure \ref{fig:water_feature} plots the
strength of water features against planet mass for
sub-Neptune planets \citep{2017AJ....154..261C}. We note a
possible dichotomy that only low-mass
($\lesssim 10~M_\oplus$) planets tend to have muted
absorption features. One explanation is that planets more
massive than 10 M have gravitational wells too strong to
allow adequate atmospheric loss, as seen in numerical
explorations of WD18.
{
  Meanwhile, due to the large optical depths in \lya
  \citep{DraineBook} and the metastable helium line
  \citep{2018ApJ...855L..11O}, a simple calculation show
  that both lines should still be observable by transmission
  spectra for planets undergoing dust outflows.} Second, the
observed $r_\p$ may differ significantly from the predicted
radius assuming a clear atmosphere
(\S\ref{sec:sim-results}). A key objective of the TESS
mission is to accurately measure the masses and radii of
$>50$ sub-Neptunes, followed by ensemble analyses of their
compositions, which may be significantly biased if leaving
dusty outflows unaccounted for. {Third, as
  $\sigma_{\rm d, ext}$ increases at shorter wavelengths,
  $\mean{r_\eff}$ in optical bands should be greater than
  infrared.  The transiting radii yielded by
  eq.~\eqref{eq:sig-ext} at $\lambda = 0.5~\micron$ are
  $\sim 10-20\%$ greater than $\lambda = 1~\micron$. Such
  phenomenon has been observed for a few exoplanets
  \citep[e.g.][]{2014A&A...570A..89E}. Extending wavelength
  coverage of transmission spectra (e.g. {\it Spitzer})
  should also be able to detect more dust-specific
  signatures.
}


\vspace*{20pt}

This work is supported by the Center for Computational
Astrophysics of Flatiron Institute, and the Department of
Astrophysical Sciences of Princeton University. We thank our
colleagues (alphabetical order): Xue-Ning Bai, Adam Burrows,
Jeremy Goodman, Xiao Hu and Kento Masuda, for helpful
discussions and comments.


\begin{thebibliography}{}
\expandafter\ifx\csname natexlab\endcsname\relax\def\natexlab#1{#1}\fi

\bibitem[{{Baines} {et~al.}(1965){Baines}, {Williams}, \&
  {Asebiomo}}]{1965MNRAS.130...63B}
{Baines}, M.~J., {Williams}, I.~P., \& {Asebiomo}, A.~S. 1965, \mnras, 130, 63

\bibitem[{{Chiang} \& {Goldreich}(1997)}]{1997ApJ...490..368C}
{Chiang}, E.~I., \& {Goldreich}, P. 1997, \apj, 490, 368

\bibitem[{{Crossfield} \& {Kreidberg}(2017)}]{2017AJ....154..261C}
{Crossfield}, I. J.~M., \& {Kreidberg}, L. 2017, \aj, 154, 261

\bibitem[{{Cubillos} {et~al.}(2017){Cubillos}, {Erkaev}, {Juvan}, {Fossati},
  {Johnstone}, {Lammer}, {Lendl}, {Odert}, \&
  {Kislyakova}}]{2017MNRAS.466.1868C}
{Cubillos}, P., {Erkaev}, N.~V., {Juvan}, I., {et~al.} 2017, \mnras, 466, 1868

\bibitem[{{Draine}(2011)}]{DraineBook}
{Draine}, B.~T. 2011, {Physics of the Interstellar and Intergalactic Medium}
  (Princeton University Press)

\bibitem[{{Draine} \& {Lee}(1984)}]{1984ApJ...285...89D}
{Draine}, B.~T., \& {Lee}, H.~M. 1984, \apj, 285, 89

\bibitem[{{Draine} \& {Malhotra}(1993)}]{1993ApJ...414..632D}
{Draine}, B.~T., \& {Malhotra}, S. 1993, \apj, 414, 632

\bibitem[{{Ehrenreich} {et~al.}(2014){Ehrenreich}, {Bonfils}, {Lovis},
  {Delfosse}, {Forveille}, {Mayor}, {Neves}, {Santos}, {Udry}, \&
  {S{\'e}gransan}}]{2014A&A...570A..89E}
{Ehrenreich}, D., {Bonfils}, X., {Lovis}, C., {et~al.} 2014, \aap, 570, A89

\bibitem[{{Fortney} {et~al.}(2013){Fortney}, {Mordasini}, {Nettelmann},
  {Kempton}, {Greene}, \& {Zahnle}}]{2013ApJ...775...80F}
{Fortney}, J.~J., {Mordasini}, C., {Nettelmann}, N., {et~al.} 2013, \apj, 775,
  80

\bibitem[{{Fossati} {et~al.}(2017){Fossati}, {Erkaev}, {Lammer}, {Cubillos},
  {Odert}, {Juvan}, {Kislyakova}, {Lendl}, {Kubyshkina}, \&
  {Bauer}}]{2017A&A...598A..90F}
{Fossati}, L., {Erkaev}, N.~V., {Lammer}, H., {et~al.} 2017, \aap, 598, A90

\bibitem[{{Fulton} {et~al.}(2017){Fulton}, {Petigura}, {Howard}, {Isaacson},
  {Marcy}, {Cargile}, {Hebb}, {Weiss}, {Johnson}, {Morton}, {Sinukoff},
  {Crossfield}, \& {Hirsch}}]{2017AJ....154..109F}
{Fulton}, B.~J., {Petigura}, E.~A., {Howard}, A.~W., {et~al.} 2017, \aj, 154,
  109

\bibitem[{{Garc{\'{\i}}a Mu{\~n}oz} \& {Cabrera}(2018)}]{2018MNRAS.473.1801G}
{Garc{\'{\i}}a Mu{\~n}oz}, A., \& {Cabrera}, J. 2018, \mnras, 473, 1801

\bibitem[{{Ginzburg} {et~al.}(2016){Ginzburg}, {Schlichting}, \&
  {Sari}}]{2016ApJ...825...29G}
{Ginzburg}, S., {Schlichting}, H.~E., \& {Sari}, R. 2016, \apj, 825, 29

\bibitem[{{Jontof-Hutter} {et~al.}(2014){Jontof-Hutter}, {Lissauer}, {Rowe}, \&
  {Fabrycky}}]{2014ApJ...785...15J}
{Jontof-Hutter}, D., {Lissauer}, J.~J., {Rowe}, J.~F., \& {Fabrycky}, D.~C.
  2014, \apj, 785, 15

\bibitem[{{Kawashima} \& {Ikoma}(2018)}]{2018ApJ...853....7K}
{Kawashima}, Y., \& {Ikoma}, M. 2018, \apj, 853, 7

\bibitem[{{Kreidberg} {et~al.}(2014){Kreidberg}, {Bean}, {D{\'e}sert},
  {Benneke}, {Deming}, {Stevenson}, {Seager}, {Berta-Thompson}, {Seifahrt}, \&
  {Homeier}}]{2014Natur.505...69K}
{Kreidberg}, L., {Bean}, J.~L., {D{\'e}sert}, J.-M., {et~al.} 2014, \nat, 505,
  69

\bibitem[{{Lammer} {et~al.}(2016){Lammer}, {Erkaev}, {Fossati}, {Juvan},
  {Odert}, {Cubillos}, {Guenther}, {Kislyakova}, {Johnstone}, {L{\"u}ftinger},
  \& {G{\"u}del}}]{2016MNRAS.461L..62L}
{Lammer}, H., {Erkaev}, N.~V., {Fossati}, L., {et~al.} 2016, \mnras, 461, L62

\bibitem[{{Li} \& {Draine}(2001)}]{2001ApJ...554..778L}
{Li}, A., \& {Draine}, B.~T. 2001, \apj, 554, 778

\bibitem[{{Lopez} \& {Fortney}(2014)}]{2014ApJ...792....1L}
{Lopez}, E.~D., \& {Fortney}, J.~J. 2014, \apj, 792, 1

\bibitem[{{Mandel} \& {Agol}(2002)}]{2002ApJ...580L.171M}
{Mandel}, K., \& {Agol}, E. 2002, \apj, 580, L171

\bibitem[{{Masuda}(2014)}]{2014ApJ...783...53M}
{Masuda}, K. 2014, \apj, 783, 53

\bibitem[{{Morley} {et~al.}(2013){Morley}, {Fortney}, {Kempton}, {Marley},
  {Visscher}, \& {Zahnle}}]{2013ApJ...775...33M}
{Morley}, C.~V., {Fortney}, J.~J., {Kempton}, E.~M.-R., {et~al.} 2013, \apj,
  775, 33

\bibitem[{{Morley} {et~al.}(2012){Morley}, {Fortney}, {Marley}, {Visscher},
  {Saumon}, \& {Leggett}}]{2012ApJ...756..172M}
{Morley}, C.~V., {Fortney}, J.~J., {Marley}, M.~S., {et~al.} 2012, \apj, 756,
  172

\bibitem[{{Murray-Clay} {et~al.}(2009){Murray-Clay}, {Chiang}, \&
  {Murray}}]{2009ApJ...693...23M}
{Murray-Clay}, R.~A., {Chiang}, E.~I., \& {Murray}, N. 2009, \apj, 693, 23

\bibitem[{{Ofir} {et~al.}(2014){Ofir}, {Dreizler}, {Zechmeister}, \&
  {Husser}}]{2014A&A...561A.103O}
{Ofir}, A., {Dreizler}, S., {Zechmeister}, M., \& {Husser}, T.-O. 2014, \aap,
  561, A103

\bibitem[{{Oklopcic} \& {Hirata}(2018)}]{2018ApJ...855L..11O}
{Oklopcic}, A., \& {Hirata}, C.~M. 2018, \apj, 855, L11

\bibitem[{{Owen} \& {Wu}(2016)}]{2016ApJ...817..107O}
{Owen}, J.~E., \& {Wu}, Y. 2016, \apj, 817, 107

\bibitem[{{Owen} \& {Wu}(2017)}]{2017ApJ...847...29O}
---. 2017, \apj, 847, 29

\bibitem[{{Parker}(1958)}]{1958ApJ...128..664P}
{Parker}, E.~N. 1958, \apj, 128, 664

\bibitem[{{Pilinski} \& {Crowley}(2015)}]{2015JGRA..120.3097P}
{Pilinski}, M.~D., \& {Crowley}, G. 2015, Journal of Geophysical Research
  (Space Physics), 120, 3097

\bibitem[{{Rafikov}(2006)}]{2006ApJ...648..666R}
{Rafikov}, R.~R. 2006, \apj, 648, 666

\bibitem[{{Ribas} {et~al.}(2005){Ribas}, {Guinan}, {G{\"u}del}, \&
  {Audard}}]{2005ApJ...622..680R}
{Ribas}, I., {Guinan}, E.~F., {G{\"u}del}, M., \& {Audard}, M. 2005, \apj, 622,
  680

\bibitem[{{Walkowicz} \& {Basri}(2013)}]{2013MNRAS.436.1883W}
{Walkowicz}, L.~M., \& {Basri}, G.~S. 2013, \mnras, 436, 1883

\bibitem[{{Wang} \& {Dai}(2018)}]{2018ApJ...860..175W}
{Wang}, L., \& {Dai}, F. 2018, \apj, 860, 175

\bibitem[{{Zhao} {et~al.}(2018){Zhao}, {Kaiser}, {Xu}, {Ablikim}, {Ahmed},
  {Evseev}, {Bashkirov}, {Azyazov}, \& {Mebel}}]{2018NatAs...2..973Z}
{Zhao}, L., {Kaiser}, R.~I., {Xu}, B., {et~al.} 2018, Nature Astronomy, 2, 973

\end{thebibliography}
\bibliographystyle{apj}

%
\end{document}